\documentclass[amsmath, amsfonts, superscriptaddress, showpacs, twocolumn, prl]{revtex4}
\usepackage{graphicx}
\usepackage{epsfig}
\usepackage{bm}
\usepackage{dcolumn}
\usepackage{amsmath}
\usepackage{amssymb}
\usepackage{color}

\DeclareFontFamily{U}{wncy}{}
\DeclareFontShape{U}{wncy}{m}{n}{<->wncyr10}{}
\DeclareSymbolFont{mcy}{U}{wncy}{m}{n}
\DeclareMathSymbol{\Sh}{\mathord}{mcy}{"58}
\DeclareMathSymbol{\Ka}{\mathord}{mcy}{"4B}

\newcommand{\be}{\begin{equation}}
\newcommand{\ee}{\end{equation}}

\begin{document}

\title{Asymmetric nanowire SQUID: linear CPR, stochastic switching, and symmetries}

\author{A.~Murphy}
\affiliation{Department of Physics, University of Illinois at
Urbana-Champaign, Urbana, Illinois 61801, USA}

\author{A.~Bezryadin} 
\affiliation{Department of Physics, University of Illinois at
Urbana-Champaign, Urbana, Illinois 61801, USA}

\begin{abstract}

We study nanodevices based on ultrathin superconducting nanowires connected in parallel to form nanowire SQUIDs. The function of the critical current versus magnetic field, $I_{C}(B)$, is multivalued, asymmetric and its maxima and minima are shifted from the usual integer and half integer flux quantum points. The nanowire interference device is qualitatively distinct from conventional SQUIDs because nanowires do not obey the same current-phase relationship (CPR) as Josephson junctions. We demonstrate that the results can be explained assuming that (i) the CPR is linear and (ii) that each wire is characterized by a sample-specific critical phase, which is usually much larger than $\pi/2$. Our proposed model offers accurate fits to $I_{C}(B)$. It explains the single-valuedness regions where only one vorticity (i.e., the order parameter winding number) is stable as well as regions where multiple vorticity values are allowed for the SQUIDs. We also observe and explain regions in which the standard deviation of the switching current is independent of the magnetic field. We develop a technique that allows a reliable detection of hidden phase-slips. Using this technique we find that our model correctly predicts the boundaries of vorticity regions, even at low currents where $I_C(B)$ is not directly measurable. 
\end{abstract}

\date{February 28 28, 2017}

\maketitle

\section{I. Introduction}
Superconducting quantum interference devices\cite{Jak,Tinkham} (SQUIDs) are known to be extremely sensitive to weak magnetic fields, and therefore various forms of superconducting loops have recently attracted significant attention\cite{McCaughan, Vasyukov, Sharon, Petkovic, Vijay,Monaco,michotte,Bourgeois, adam}. Nanowire networks\cite{Pann} and loops\cite{Hopkins-Science, Gurtovoi, Belkin, Arpaia, Burlakov-JETP2014, Burlakov-JETP2011, voss, pekker} are qualitatively distinct from conventional Josephson junction and SQUIDS due to the linear nature of the nanowire CPR \cite{Arpaia}. In contrast, conventional Josephson junctions obey a sinusoidal CPR. Nanowire SQUIDs have been used in important applications such as the detection of macroscopic quantum tunneling in magnetic systems with large spins\cite{wernsdrfer}. 

The properties of unshunted conventional SQUIDs composed of two Josephson junctions are well known \cite{Tinkham}. In the simplest case where the loop inductance is negligible, the critical current of the SQUID is a periodic, single-valued function of the magnetic field, and its maxima correspond to integer multiples of the flux quantum, while its minima occur at half flux quantum plus an integer number of flux quanta\cite{Tinkham}. If the SQUID is asymmetric, i.e. if the critical currents of the two branches forming the SQUID are different, then the conditions listed above remain true, but the critical current modulation does not go all the way to zero at half flux quantum. The fact that the maxima and minima of the $I_{C}(B)$ function coincide with integer and half-integer normalized flux values is due to the sinusoidal nature of the CPR for each weak link and the associated critical phase of $\pi/2$.

Here we present experiments and propose a model for asymmetric nanowire SQUIDs in which the critical phase is significantly larger than $\pi$/2. This fact leads to the occurrence of multiple metastable states which differ by their winding number (vorticity). We elucidate the qualitative changes which the function $I_{C}(B)$ exhibits in such cases. The devices exhibit a number of characteristics which make them qualitatively different from the conventional unshunted SQUIDs \cite{Tinkham}. We observe that the critical current of the nanowire SQUID is multi-valued and its minima and maxima can shift strongly from the usual integer and half-integer flux quanta values. These features have been observed previously \cite{podd, Gurtovoi, sivakov, hasselbach, harza}. We also observe that the general shape of the $I_{C}(B)$ curve is made of linear segments rather than being sinusoidal. At temperatures much lower than $T_C$, a linear CPR (and therefore a linear relationship between critical current and field) is both expected \cite{likharev, Bardeen, Tinkham, wei, bagwell} and has been observed experimentally \cite{Arpaia, hasselbach,Burlakov-JETP2014, sivakov, harza}. Advanced computational simulations based on Ginzburg-Landau theory, which is known to be valid at temperatures near the critical temperature $T_C$ \cite{Meidan}, have been used to simulate the critical current versus field dependence of nanowire loops previously \cite{podd, hasselbach, sivakov}. While some of these computational models do calculate a piecewise linear or near-linear dependence of critical current on magnetic field \cite{sivakov, hasselbach}, these authors do not present theoretical analysis of cases in which the loop is asymmetric. Additionally, none plot theory on top of experimental data to allow direct comparison of the two. Here we propose a simple model, based on a linear CPR, which allows accurate fitting of the critical current versus field dependence. The model predicts the multi-valuedness, shifts in maxima and minima, and the linearity of the $I_C (B)$ function. Thus we confirm Bardeen's prediction that the CPR of a thin wire is linear at very low temperatures \cite{Bardeen}. We address asymmetric systems quantitatively and plot theoretical curves on top of experimental data for direct comparison. 

We also observe and discuss unusual plateaus in the standard deviation of the switching current distribution. Furthermore, we observe that the regions of stability of the vorticity, namely the Little-Parks diamonds \cite{sivakov, Little, Parks}, can overlap significantly, thus generating multivaluedness of the critical current and of the vorticity at a fixed magnetic field. Yet we find some magnetic field - bias current parameter regions, which we call unique-vorticity diamonds, in which only one vorticity is stable. These results open doors to vorticity-manipulation experiments. We observe that the critical current versus field function $I_C(B)$ is symmetric with respect to the origin, if both positive and negative branches are included. This fact is explained within our linear-CPR model of a nanowire SQUID. Finally, we observe the presence of hidden phase-slips, i.e. phase slips which are not accompanied by the switching of the device to the normal state, as predicted in our model. 

\section{II. Experiment}

\begin{figure}[t!]
  \includegraphics[width=6cm]{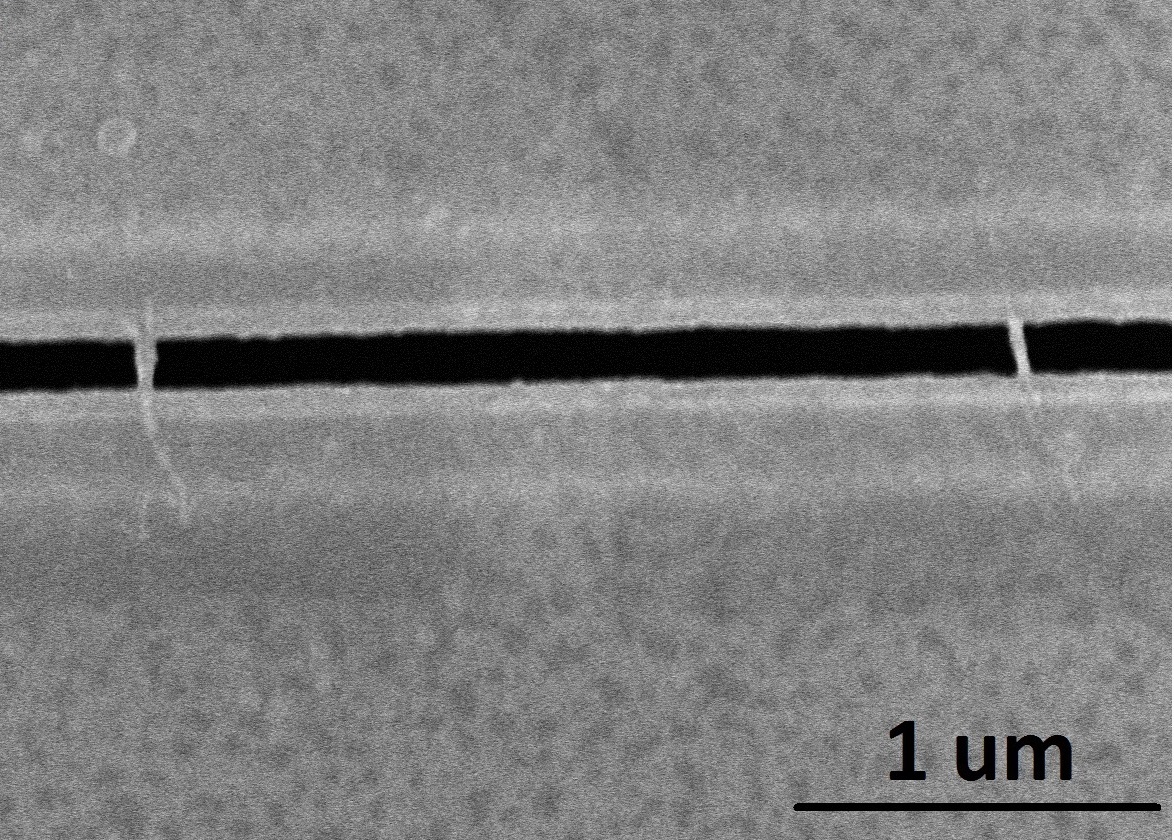}
  \caption{An SEM image of Device 7715s1. Two nanowires (gray) lay across a 140 nm wide trench (black). The distance between the wires is 2.5 $\mu$m. The superconducting electrodes appear as the gray areas above and below the nanowires.   
 }\label{Fig-sem}\vskip-.5cm
\end{figure}

All three measured nanowire SQUIDs, Device 7715s1 (Fig. \ref{Fig-sem}), Device 51215s3 and Device 31414s1 are produced by a molecular templating method\cite{Bezryadin, Bezryadin-JP}. In brief, the nanowires were made by depositing carbon nanotubes across a 140 nm trench on a Si chip coated with a bilayer of SiO$_2$ and SiN. A layer of Mo$_{25}$Ge$_{75}$ was sputtered on the entire chip coating both the carbon nanotubes and the SiN surface. This layer was 18 nm thick for Device 7715s1, 17 nm thick for Device 51215s3 and 10 nm thick for Device 31414s1. This process creates both the Mo$_{25}$Ge$_{75}$ nanowires and the wide electrodes connected to them simultaneously, thus contact resistance does not occur. Contact pads and electrodes were then patterned by photolithography such that after etching with H$_2$O$_2$, only the two desired nanowires remained as weak superconducting links between the electrodes. The width of the electrodes is 20 $\mu$m. 

Each device consists of two nonidentical nanowires which are connected in parallel to the two electrodes, forming an asymmetric superconducting loop (Fig. \ref{Fig-sem}). The bias current flows from one of these electrodes, through the pair of nanowires, to the second electrode. One nanowire of Device 7715s1 is 42 nm wide and 140 nm in length, and the other one is 26 nm wide and 158 nm in length. The nanowires are separated by 2.5 $\mu$m. Device 51215s3 consists of a 29 nm wide by 190 nm long nanowire separated from a 19 nm wide by 170 nm long nanowire by distance 1.3 $\mu$m. Device 31414s1 consists of a 35 nm wide by 225 nm long wire separated from a 23 nm wide by 216 nm long wire by a distance of 2.6 $\mu$m.

Device 7715s1 is measured at 320 mK in a He3 system, and all devices are measured at and above 1.5 K in a He4 cryostat. A current biasing is achieved by placing larger resistors in series with the device and a function generator. Two resistors have been used, 1 k$\Omega$ and 47 k$\Omega$. Current is calculated using Ohm's law by measuring the voltage across the 1 k$\Omega$ resistor whereas the voltage on the sample is measured on the contact pads, which are connected to the electrodes. A sinusoidal bias current with a sweep frequency of 1.1 Hz is applied to Device 7715s1 in the He3 measurements, while a frequency of 3.5 Hz (Device 7715s1 and Device 51215s3) or 11 Hz (Device 31414s1) is applied for measurements of critical current performed in the He4 setup.

At temperatures sufficiently below $T_C$, as the current is increased from zero, the voltage across the device is initially zero, but at some critical value of the applied current the voltage suddenly jumps from zero to a large value of the order of tens of mV. Such a sudden jump indicates that the device has become normal. The current at which this transition takes place is recorded as the critical current  $I_{C}$.

\begin{figure}[t!]
  \includegraphics[width=8cm]{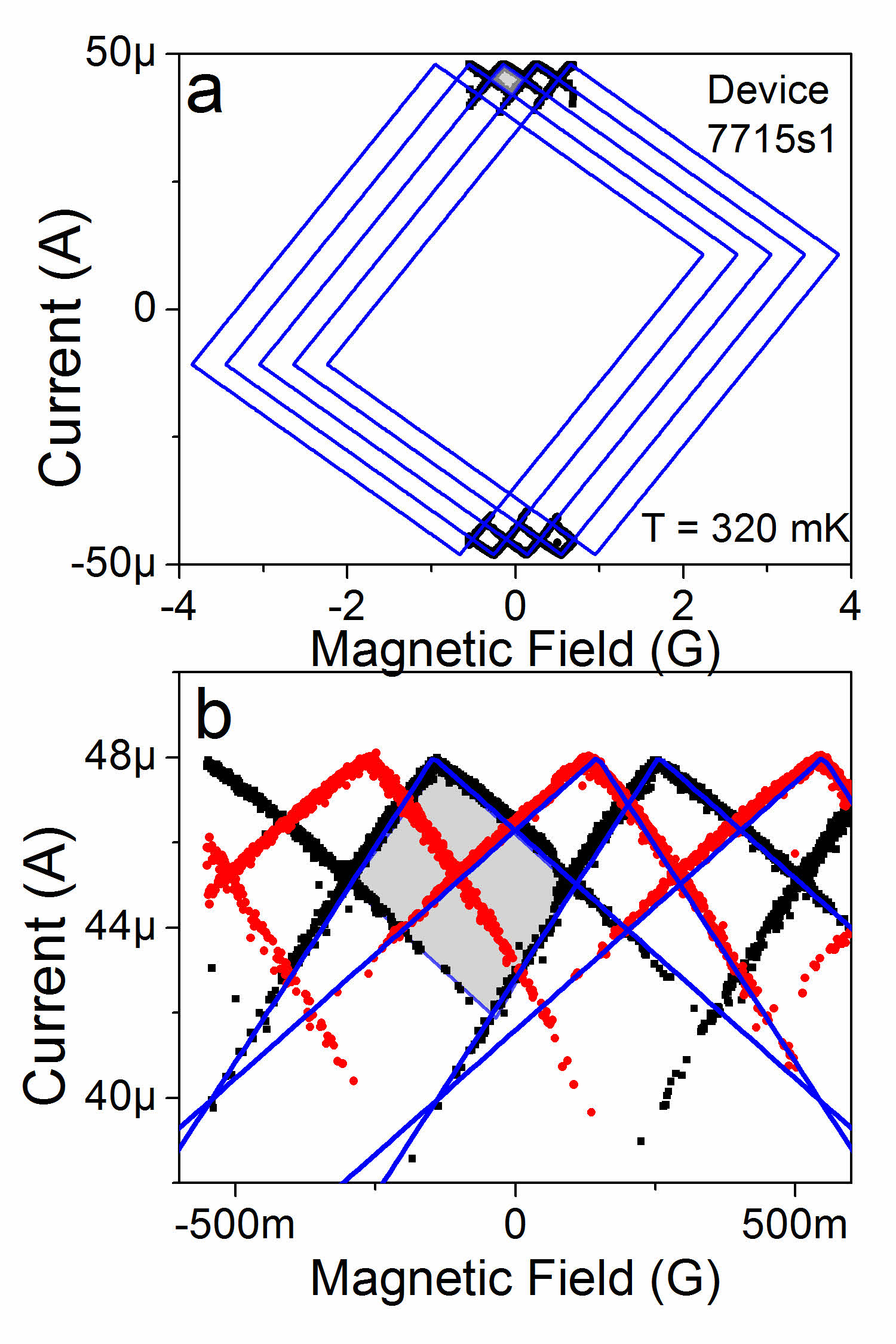}
  \caption{a) The critical current (black points) plotted against magnetic field. Fits, in solid lines forming diamond shapes, show the Little-Parks diamonds generated by our model (see text). They predict the critical currents associated with states characterized by fixed vorticity values. From left to right, the vorticity of the fitting curves increments from n$_v$ = -2  to 2. Fit parameters are listed in Table \ref{Table}. b) For clarity, the absolute value of the critical current and fits of the critical current for states n$_v$ = 0 and 1 are plotted. Black squares denote positive critical currents and red circles denote the absolute value of the negative critical currents. The shaded area in both figures shows the positive-current unique-vorticity diamond of the vorticity state n$_v$ = 0.
 }\label{Fig-isw}\vskip-.5cm
\end{figure}

In Figure \ref{Fig-isw}a we plot the critical current (black dots) of the SQUID 7715s1 versus magnetic field $B$. The measured critical current is a multivalued periodic function of magnetic field, composed of approximately linear segments and resembling a periodic sequence of diamonds. A linear dependence of the critical current on the magnetic field has been observed previously \cite{Arpaia, hasselbach, sivakov,Burlakov-JETP2014, harza}. In Fig. \ref{Fig-isw}b, the positive critical current $I_{C+}$ is compared to the negative critical current $I_{C-}$, which was multiplied by $-1$. In conventional SQUIDs the maximum in the switching current always occurs at zero field \cite{Tinkham} and the minimum occurs if the applied flux equals half of the flux quantum. However this is not true in general in our two-wire devices. In this example of the representative device 7715s1, measured at $T=$320 mK (Fig.\ref{Fig-isw}b), once can see that $I_{C+}$ = $I_{C-}$ at $B=0$. Yet, neither the maxima nor the minima occur at $B=0$.

\section{III. Model}
The results can be understood as follows. The total bias current is split between the two wires. We assume the current in each individual wire is given by a linear current-phase relationship

\begin{equation}
I_j = I_{C,j} \phi_j / \phi_{C,j}
\end{equation}

where $j$ = 1 or 2 is the wire number. $I_j$ is the supercurrent through the wire $j$, $I_{C,j}$ is the critical current of wire $j$, $\phi_j$ is the difference of the phase of the complex superconducting order parameter taken between the end points of the wire $j$. We also introduce the notion of the critical phase difference $\phi_{C,j}$ ($j$ is the wire number) which is the phase difference at which the supercurrent reaches its maximum possible value and the superconductivity gets destroyed. The critical phase is proportional to the wire length and $\phi_C>\pi/2$ ($\phi_C=\pi/2$ if the wire is much shorter than the coherence length which is about 7 nm in our samples). 

The second key component to build this quantitative model is the fact that the order parameter must be single-valued on the closed loop. The total phase around the superconducting loop must be an integer multiple of $2\pi$ due to the single-valuedness of the order parameter. Therefore the phases across each wire and the electrodes must add as \cite{Hopkins-Science}

\begin{equation}
\phi_1 - \phi_2 + 2 \delta  = 2 \pi n_v 
\end{equation}

Here, the vorticity (the winding number) of the SQUID loop is $n_v$. The phase difference within each electrode $\delta = \delta (B)$ (defined between the ends of the two nanowires connected to the same electrode) is assumed to be the same on both the electrodes; thus the factor of 2 occurs in the phase balance equation given above. This Meissner phase difference is $2\delta (B)=2\pi (B/\Delta B)$, where $\Delta B$ is the Little-Parks period and $B$ is the external field applied perpendicular to the SQUID loop \cite{Hopkins-Science}. In this simplified model only the kinetic energy of the superconducting condensate is taken into account and the magnetic field distortion by the Meissner effect is neglected. A more rigorous theoretical model might need to include the magnetic moment of the supercurrent in the loop and its interaction with the applied field \cite{nikulov, gurtovoi-comment}. In the present model, the phase gradients in the electrodes are assumed to be created by the Meissner current in the electrodes and negligibly affected by the applied supercurrents through the nanowires because the wires are very weak superconductors compared to the electrodes \cite{Hopkins-Science}. 

Combining equations 1 and 2, and the requirement that superconductivity should be destroyed if $\phi_{j} \geqslant \phi_{C,j}$, we have calculated the total critical current of the nanowire SQUID for a given vorticity $n_v$ and magnetic field B. We assume the total critical current of the device, $I_C(B)$, equals the smallest total applied current at which the current across either wire reaches its critical value.  Different values of $n_v$ result in different critical currents. When the critical current of a vorticity state $n_v$ is plotted against magnetic field, we find the boundaries of the region in which the vorticity state $n_v$ is stable. We call the region in which a given vorticity is stable a Little-Parks diamond. Outside its Little-Parks diamond, the vorticity state $n_v$ cannot exist because the critical current of at least one wire will be exceeded. When the system reaches the boundary of its vorticity state, either the vorticity of the system will change to a new, stable vorticity providing a larger critical current, or the device will switch to the normal state.

The critical currents of Device 7715s1 are calculated using equations 1 and 2 and shown in Fig. \ref{Fig-isw} by the straight lines. The fitting parameters are listed in Table I. The model gives good fits, thus confirming that the CPR is linear with a high accuracy. Many Little-Parks diamonds overlap over wide ranges of fields and currents. This explains why the critical current is multivalued. Another interesting fact is that the model does not involve the geometric inductance of the nanowires. Thus only their kinetic inductance is essential. Therefore it should be possible to reduce the dimensions of such nanowire SQUIDS by a large factor without compromising their performance (since the kinetic inductance can be large even if the wire is small, if the critical current of the wire is small).

A linear CPR has been predicted for thin wires at $T = 0$ \cite{Bardeen, wei, bagwell, likharev, Tinkham}. Additionally, many computationally advanced models based on Ginzburg-Landau theory have been developed \cite{sivakov, hasselbach, podd}. Here, we have shown that a simple model based on a linear CPR provides excellent fits to our critical current data. Below, we will show how this model leads to a thorough understanding of the critical current vs magnetic field function, reveals the process by which the system switches from the superconducting state to the normal state, and how it correctly predicts the existence of hidden phase-slips.

\section{IV. Analysis}

According to our model, if the wires are different, the optimal vorticities (the one which produces the largest $I_C(B)$) are not always equal for positive and negative currents. At the largest currents at which the sample is still superconducting, only the optimal vorticity state is stable because the critical currents of all other vorticity states have been exceeded. The region in field and current in which the optimal vorticity is the only stable superconducting state will be referred to as the unique-vorticity diamond. For example, the unique-vorticity diamond for state $n_v$ = 0 at positive currents of Device 7715s1 is shown as the shaded region in Fig.~\ref{Fig-isw}. If the system is superconducting within the unique vorticity diamond then the vorticity state is known. This suggests that nanowire SQUIDs may be applicable as memory devices  using the unique-vorticity diamond to write a known vorticity state. At low current bias, the device can have many different vorticity values which are metastable. This metastability results in a multivaluedness of the critical current. It should be noted that the model predicts the existence of many critical currents for a fixed field. Yet, experimentally, we cannot see all of them simply by measuring the critical current. This fact indicates that when the bias current reaches the critical current for a given vorticity state, the system is sometimes able to modify its vorticity without switching to the normal state. 

Each maximum in magnitude of $I_C(B)$ and the two critical current branches extending from it correspond the critical currents associated with a particular vorticity state. We define a critical current branch as a continuous line segment of critical current when plotted versus magnetic field. The maximum itself occurs at the field when both wires reach their individual sample-specific critical currents (and critical phases) simultaneously. The reason that the crossing branches have different slopes is due to the fact that they represent different wires reaching their corresponding critical current and critical phase.

\begin{figure}[t!]
  \includegraphics[width=8.5cm]{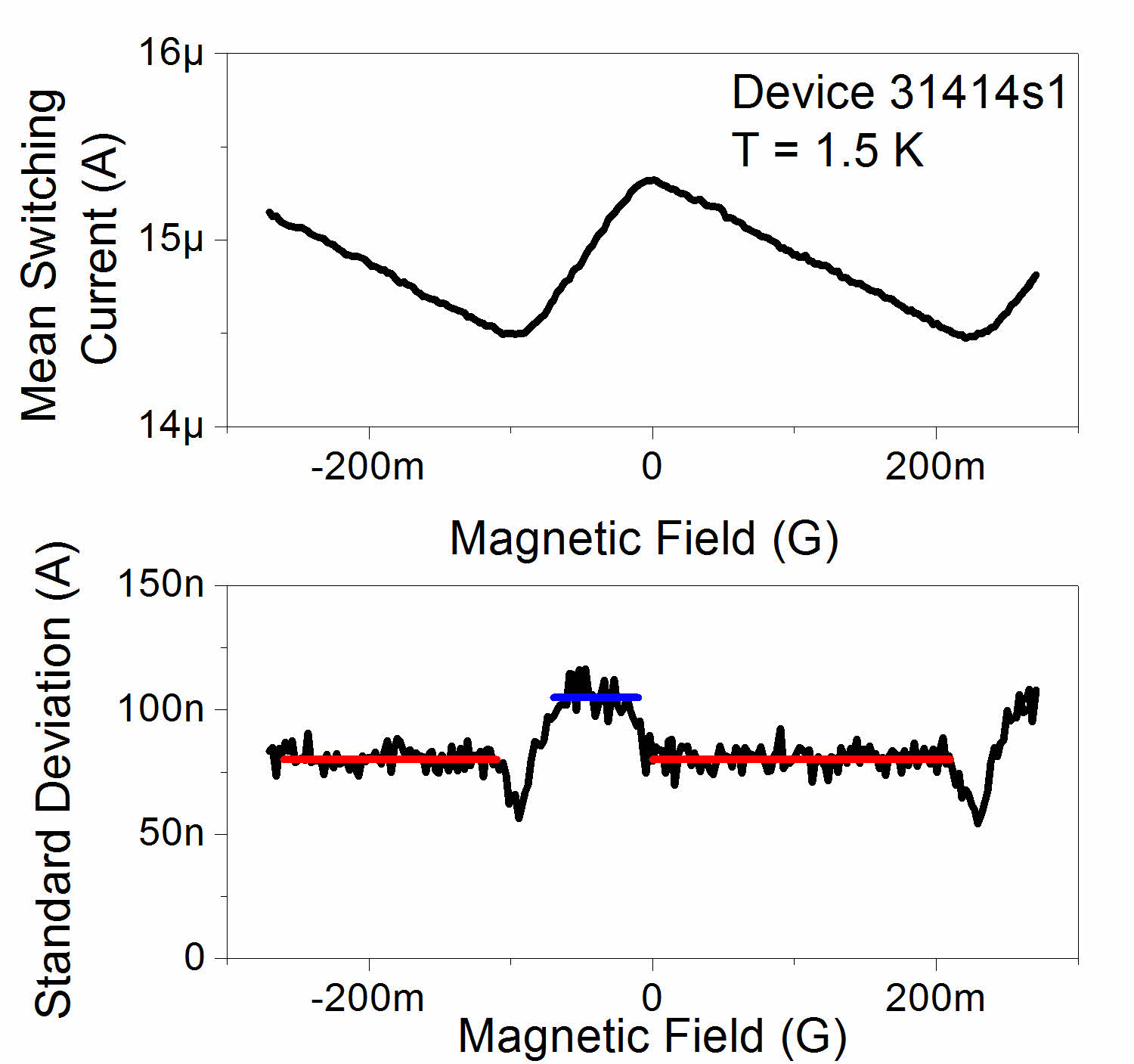}
  \caption{a) The mean of the switching current distribution vs magnetic field  for Device 31414s1. Device 31414s1 only shows one switching current branch at each magnetic field. b) The standard deviation of the switching distribution vs magnetic field is a periodic sequence of plateaus. Along each plateau, a switching event is caused by a particular wire reaching its critical current. Horizontal lines in red (at 80 nA) and blue (at 105 nA) have been plotted for reference.
 }\label{Fig-sigma}\vskip-.5cm
\end{figure}

\textit{Standard Deviation:} In measurements on superconducting junctions in which the current is slowly increased from zero, the measured critical current, often called the switching, is typically slightly less than the depairing current of the wire \cite{kurkijarvi}. Thermal or quantum fluctuations cause the system to escape from the superconducting state before the depairing current is reached. This creates a stochastic distribution of switching current measurements.  In our model above we have treated the average switching current and theoretical critical current as if they are equal, for simplicity. We plot the average and standard deviation of the switching current distributions of Device 31414s1 at 1.5 K in Fig.~\ref{Fig-sigma}. We choose to analyze the switching current distribution of Device 31414s1 because it shows only one critical current branch at any magnetic field. The switching current is measured 300 times at each magnetic field. The average switching current is plotted in Fig.~\ref{Fig-sigma}a and the standard deviation $\sigma$ of the switching current distribution is shown in Fig.~\ref{Fig-sigma}b. The standard deviation of the switching current distribution is a periodic sequence of plateaus. Along each branch of the critical current, the standard deviation in constant. This fact gives us an important insight into the switching mechanism. As we will discuss below, these plateaus indicate that the switching originates from a single wire in which the critical condition is reached first. At the minimum of the average switching current, there is a dip in the standard deviation. 

In order to understand the plots in Fig.~\ref{Fig-sigma}, consider only the cases in which a switching event occurs because the total current in wire 1 $I_1$ reaches its critical value $I_{C,1}$ and the system is in some vorticity state $n_v$. The total applied current when $I_1 = I_{C,1}$, i.e., the depairing current, can be calculated using Eq. 1, 2, and conservation of current. The depairing current changes linearly with magnetic field. However, as long as a switching event is caused by wire 1 reaching its critical current, $I_1$ will equal $I_{C,1}$ at the depairing current. Note that the total current in wire 2 may be sub-critical. The standard deviation of the switching current distribution is related to the rate of switching events as a function of the total current in wire 1. This rate does not change noticeably with magnetic field because when the switching happens the total current in wire 1 is near its critical current, at any value of magnetic field. Note that we still only consider that case where a switching event is cause by wire 1 reaching its critical current. Thus, as long as a switching event is caused by wire 1 reaching its switching current, $\sigma$ is constant. Similarly, $\sigma$ will be constant in the range of magnetic fields in which switching events are caused by wire 2 reaching its critical current.

Each plateau in Fig.~\ref{Fig-sigma}b is a region in magnetic field over which a switching event is caused by on particular wire reaching its critical current. However, a periodic dip in standard deviation occurs at fields where the switching current is a minimum. We speculate that this dip may be due to coincidences of multiple phase-slips. A phase-slip is a process by which a vortex crosses one of the wires, causing the vorticity of the loop to change by $\pm$1 \cite{little}. If a non-zero current flows in the wire, the phase-slip will dissipate heat. If enough heat is dissipated such that the depairing current of the device is reduced below the applied current, the system will switch to the normal state. If a single phase-slip does not produce enough heat to drive the system normal, then a coincidence of simultaneous phase-slips may be necessary to switch the device to the normal state \cite{sahu, shah}. From our model, we know in the region just below the minimum in the switching current, two vorticity states are metastable (see, for example, Fig.~\ref{Fig-isw}b). Thus, it is possible that a single phase-slip will change the vorticity of the loop without driving the device normal. It has been shown on single nanowire devices that when coincidences of multiple phase-slips are responsible for driving the system normal, standard deviation decreases \cite{sahu}.

\begin{figure}[t!]
  \includegraphics[width=8.5cm]{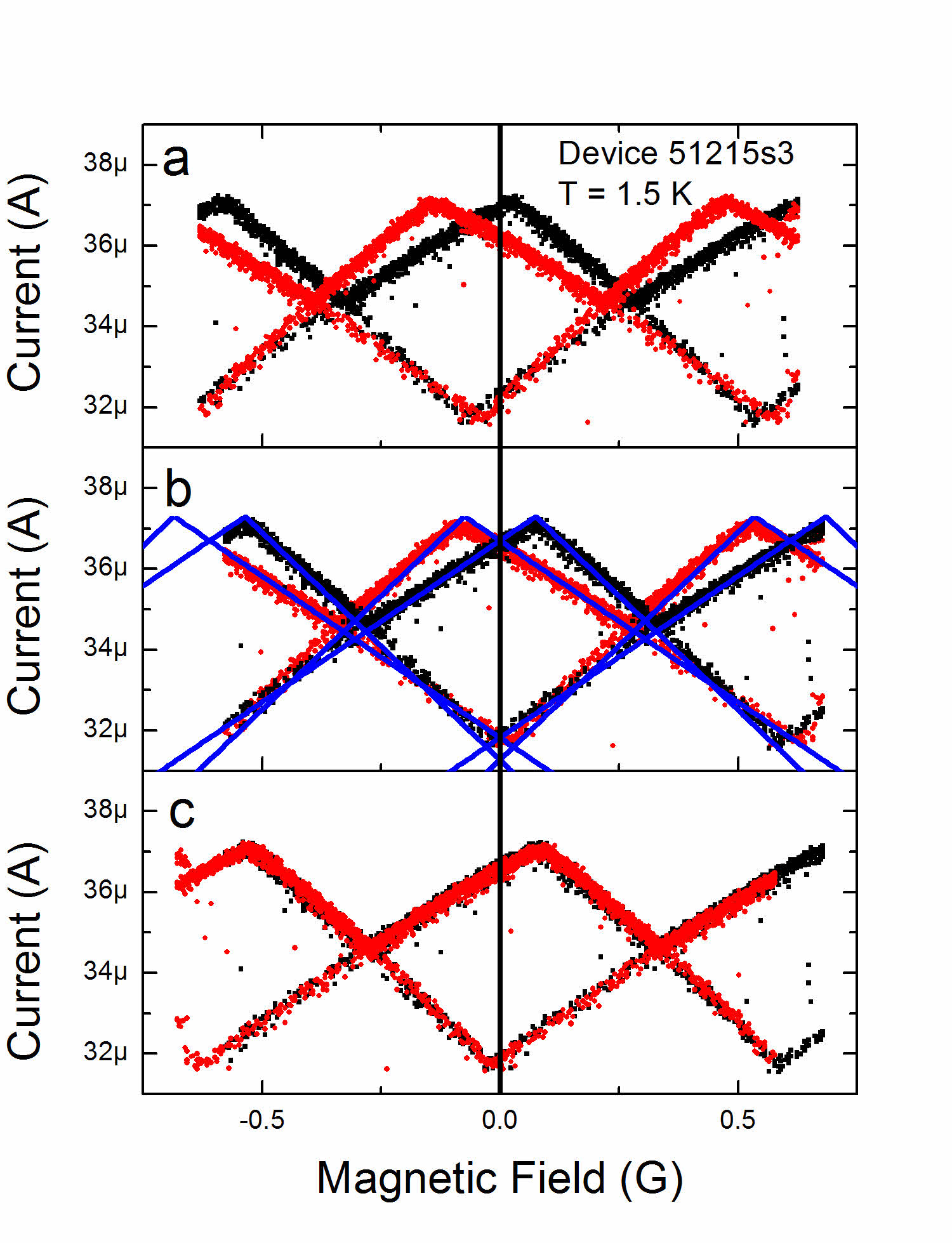}
  \caption{a) Raw critical current data. The magnitudes of the positive critical current (black squares) and negative critical current (red circles) are not equal at zero field (black line). b) The critical current is shifted with respect to magnetic field such that the largest magnitude positive and negative critical currents are equal at zero field. This data is fit to our model and fits are shown in blue lines. Fit parameters are listed in Table \ref{Table}. c) The magnitude of the negative critical current is flipped about B = 0. The overlap of the black and red curves is consistent with the physical symmetry of the system: Reversing the direction of the current and the applied field at the same time should produce no change to the system.
 }\label{Fig-iswflip}\vskip-.5cm
\end{figure}

\textit{Symmetry 1}: A transformation consisting of reversing both the direction of applied current and the direction of the magnetic field will reproduce the initial untransformed state \cite{Burlakov-JETP2014}. We illustrate the symmetry transformation on Device 51215s3, in Fig. \ref{Fig-iswflip}. In Fig.~\ref{Fig-iswflip}a the raw critical current data is presented. At zero vorticity and zero field, we expect the magnitudes of $I_{C+}$ (black squares) and $I_{C-}$ (red dots) to be equal.  Therefore in Fig.~\ref{Fig-iswflip}b we shift the data with respect to zero field (the offset may be due to the Earth field or some other unaccounted sources). A positive (black) and negative (red) branch of critical current (corresponding to the same vorticity $n_v = 0$) now intersect at $B=0$ (Fig. \ref{Fig-iswflip}b). The fits (blue lines) produced by our model exhibit an excellent agreement with the data. In Fig.~\ref{Fig-iswflip}c the symmetry transformation is completed by multiplying by -1 the magnetic field value of each data point corresponding to the negative critical currents. Thus modified negative branches (red) match perfectly well the corresponding positive branches (black) (see Fig.~\ref{Fig-iswflip}c). To summarize, the discussed current-field reversal symmetry can be expressed as $-I_{C-}(-B)=I_{C+}(B)$ where the critical current function is the complete multivalued function.

\textit{Symmetry 2}: At integer flux quanta and half-integer flux quanta the positive critical current of some vorticity state $n_{v+}$ and the negative critical current of the matching vorticity state $n_{v-}$ are equal to each other in magnitude, i.e., $I_{C+}(B,n_{v+}) = -I_{C-}(B,n_{v-})$. For this to hold true, the vorticity states $n_{v+}$ and $n_{v-}$ must be related by $n_{v+} + n_{v-} = \mathrm{int}(\frac{2B}{\Delta B})=\mathrm{int}(2\delta(B)/\pi)$ where $\Delta B$ is the period. Mathematically, magnetic fields of integer flux quanta can be expressed as $B=n\Delta B$ and half-integer flux quanta magnetic fields can be expressed as $B=(n+1/2)\Delta B$. Here $n$ is an integer.  The experiment supports our conclusions: The magnitudes of the negative (red) and the positive (black) experimental critical current curves in Fig.~\ref{Fig-isw}c intersect at integer and half-integer flux quanta, i.e., at $B=n\Delta B$ and $B=(n+1/2)\Delta B$.  To derive this equality theoretically, we will consider the cases when the current through wire $j=1$ reaches its positive critical value $I_1 = I_{C,1}$ and when the current through wire $j=1$ reaches its negative critical value $I_1 = -I_{C,1}$. In these cases, the total current $I_{Total}$ through the device (which is the total critical current of the device) can be written as

\begin{equation}
I_{Total} = I_{C+}(B,n_{v+}) = I_1 + I_2 = I_{C,1} + I_2
\end{equation}

for positive current and

\begin{equation}
I_{Total} = I_{C-}(B,n_{v-}) = I_1 + I_2 = -I_{C,1} + I_2
\end{equation}
 
for negative current. Here, $n_{v+}$ and $n_{v-}$ are two unknown vorticity states. 

We can then calculate the current in wire 2 I$_2$ using Equations 1 and 2. At positive currents

\begin{equation}
 I_2 = \frac{I_{C,2}}{\phi_{C,2}}(\phi_{C,1} + 2 \delta - 2 \pi n_{v+})
\end{equation}

and at negative currents

\begin{equation}
 I_2 = \frac{I_{C,2}}{\phi_{C,2}}(-\phi_{C,1} + 2 \delta - 2 \pi n_{v-}).
\end{equation}

Next, we multiply the negative total critical current by -1 and set it equal to the total positive critical current.

\begin{equation}
\begin{split}
 I_{C,1} + \frac{I_{C,2}}{\phi_{C,2}}(\phi_{C,1} + 2 \delta - 2 \pi n_{v+}) = \\
-(-I_{C,1} +  \frac{I_{C,2}}{\phi_{C,2}}(-\phi_{C,1} + 2 \delta - 2 \pi n_{v-}))
\end{split}
\end{equation}

This reduces to

\begin{equation}
 2 \delta - 2 \pi n_{v+} = - 2 \delta + 2 \pi n_{v-}
\end{equation}

or more simply,

\begin{equation}
n_{v+} + n_{v-} =  \frac{2}{\pi}\delta.
\end{equation}

Recall that $\delta$, the phase difference between the ends of wire 1 and wire 2 within an electrode and is related to the magnetic field by $\delta = \pi B / \Delta B$ where $\Delta B$ is the period. Thus we find that

\begin{equation}
n_{v+} + n_{v-} = \frac{2B}{\Delta B}.
\end{equation}

For this equation to hold, $\frac{2B}{\Delta B}$ must be an integer (as the vorticity values must also be integers). Therefore, solutions can be found when $B = n \Delta B / 2$ where n is an integer, i.e., when the magnetic field is either at an integer multiple of the period (and the flux is at an integer flux quanta) or when the magnetic field is at an integer plus one-half period (and the flux is at a half-integer value of flux quanta).

This analysis considered positive total critical currents achieved when $I_1 = I_{C,1}$ and the total negative critical current when $I_1 = -I_{C,1}$. A similar analysis can be done considering the total positive critical current when $I_2 = I_{C,2}$ and the total negative critical current when $I_2 = -I_{C,2}$. The same result will be found. We therefore conclude that at integer flux quanta and half-integer flux quanta, for each positive critical current, there will be a negative critical current of equal magnitude and vice versa. This relationship can be written as 

\begin{equation}
I_{C+}(B,n_{v+}) = - I_{C-}(B,n_{v-})
\end{equation}

where $ n_{v+} + n_{v-} = \mathrm{int}(\frac{2B}{\Delta B})$. In the general condition given above, the integer vorticity values $n_{v+}$ and $n_{v-}$, which index the positive and the negative critical current branches respectively, may be equal or not equal. For example, at zero field $I_{C+}(B =0,n_{v+}=0) = -I_{C-}(B =0,n_{v-}=0)$ at the largest magnitude critical currents, but $I_{C+}(B=0,n_{v+}=1) = -I_{C-}(B=0,n_{v-}=-1)$ is also true at a smaller magnitude critical current. 

The maxima in the critical current deviates from zero field if the critical phases of the two wires are different. The maximum in $I_C (B)$ at zero vorticity occurs when the total critical current is equal to the sum of the individual critical currents of each wire, $I_{C} = I_{C,1} + I_{C,2}$, and the phases across both wires have reached their critical phases simultaneously. Equation 2 becomes $\delta(B) = \frac{1}{2} (\phi_{C,2} - \phi_{C,1})$ which is nonzero as long as the two critical phases are not equal. Recall that $\delta(B)$ is linearly proportional to B \cite{Hopkins-Science}. 

Using our model, we can calculate the conditions for which multiple vorticity states are metastable. For example, at zero field and zero current, if we assume the two wires are identical ($I_{C,1}=I_{C,2}$ and $\phi_{C,1}=\phi_{C,2}$) then $\phi_1 = - \phi_2$ (this is found using Eq. 1 and conservation of current) and Eq. 2 becomes

\begin{equation}
2 \phi_1 = 2 \pi n_v.
\end{equation}

The value of $\phi_1$ can range from $-\phi_{C,1}$ to $\phi_{C,1}$. Thus, the value of $n_v$ at zero current and field can be any integer between $-\phi_{C,1}/\pi$ and $\phi_{C,1}/\pi$. So, as long as $\phi_{C,1}, \phi_{C,2}  \geq \pi$, vorticity states $n_v = -1$, $0$ and $1$ will be stable at zero field and current. It has been shown that asymmetric SQUIDS composed of Josephson junctions can have multiple metastable vorticity states \cite{duzer}, however, this result is derived considering geometric inductance. Our nanowire loops depend on kinetic inductance rather than geometric inductance, making our analysis of the metastability of vorticity states qualitatively different from studies of traditional Josephson junction SQUIDs. Note that this analysis is performed assuming the CPR is linear which is not expected to be true for very short nanowires.

\begin{figure}[t!]
  \includegraphics[width=8.5cm]{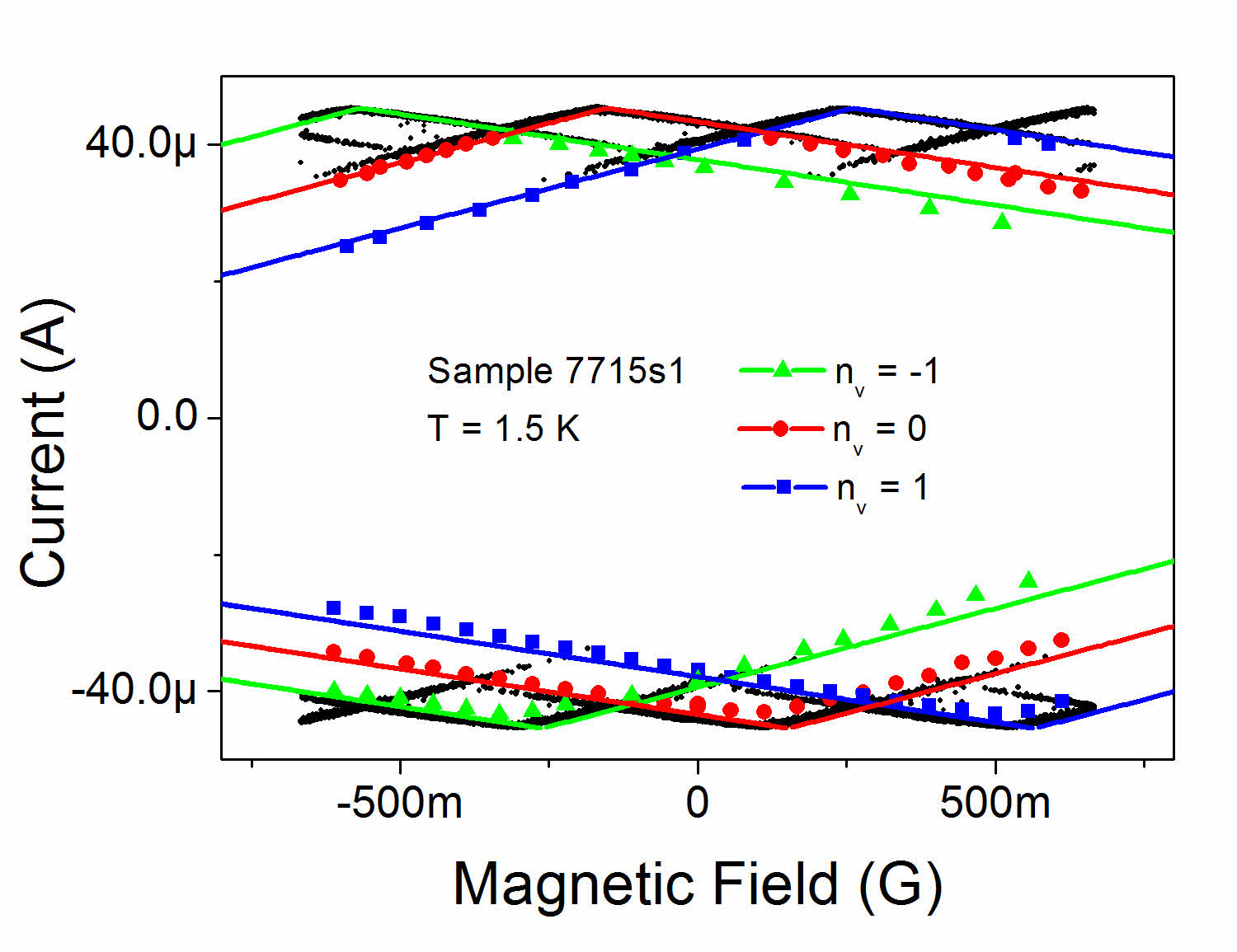}
  \caption{The borders of vorticity states $n_v = -1$, $0$ and $1$ for Device 7715s1 are found experimentally as described in the text and are plotted as green, red and blue points. Solid lines show theoretical fits. Black points show the switching currents. 
 }\label{Fig-hiddenps}\vskip-.5cm
\end{figure}

\textit{Hidden phase-slips} Next, we demonstrate a method by which we can observe phase-slips which do not produce switching events. We begin with the Device 7715s1 in the superconducting state and in the unique vorticity diamond for state $n_v = 0$. We drive the system to a value of current and field, which we call a testing point, and then return the system to the unique vorticity diamond of state $n_v=0$. If at any point, the system switches to the normal state (either upon reaching the testing point or returning to the unique vorticity diamond, we record the testing point as a boundary of state $n_v=0$. If a phase-slip at a boundary of the vorticity state produces enough heat to switch the system to the normal state, we will immediately observe a switching event. If a phase-slip at a boundary does not produce enough heat to drive the system normal, then the system must switch to a new metastable vorticity state. It must then cross the boundary for this new vorticity state upon returning to the unique vorticity diamond for state $n_v=0$. This boundary crossing results in a phase-slip which may drive the system normal. If no switching event is observed, then the testing point is incremented farther away from the unique vorticity diamond. This process is repeated for states $n_v = -1$ and $n_v = 1$. Results are shown in Fig.~\ref{Fig-hiddenps}. Data was taken in a $^4$He cryostat at 1.5 K. Black points are the critical currents. Green, red and blue points are the testing points which correspond to the borders of vorticity states $n_v = -1, 0$ and $1$. Solid lines show theoretical fits and fitting parameters are listed in Table \ref{Table}. We find that the boundaries of each vorticity states matches theoretical predictions. We are able to observe signatures of phase-slips which do not produce switching events (at low currents) and confirm the accuracy of our model.

\begin{table}[]
\centering
\caption{Fitting Parameters}
\label{Table}
\begin{tabular}{ | l | l | l | l | l |}
\hline
Device  & I$_{C1}$ ($\mu$A) & I$_{C2}$ ($\mu$A) & $\phi_{C1}$ (rad) & $\phi_{C2}$ (rad)\\
\hline

7715s1 (at T = 0.3 K) & 16.9              & 31.1              & 23.6        & 21.1        \\
7715s1 (at T = 1.5 K) & 15.9 & 29.5 & 21.9 & 19.4 \\
51215s3 & 21.5              & 15.8              & 19.6        & 20.4     \\
31414s1 & 10.0 & 5.3 &19.8 & 23.9\\
\hline
\end{tabular}
\end{table}

\section{V. Conclusions}

In conclusion, nanowire SQUIDs are qualitatively different from conventional SQUIDs because the critical phase of the nanowires involved is much larger than $\pi/2$. The critical current is multivalued. At integer flux quanta and half-integer flux quanta, the magnitudes of the positive and negative critical currents are equal, but are not necessarily maxima or minima. The stability regions of vorticity states are described by Little-Parks diamonds, and the critical current function is composed of linear segments. We find that a linear segment in the critical current vs magnetic field function corresponds to a plateau in the standard deviation in the switching current distribution. A single line segment / plateau corresponds to the situation where the current in one wire reaches its critical current for some vorticity state. We test our model by detecting hidden phase-slips which do not produce switching events along predicted boundaries of vorticity states.

We thank D. Averin and J. Ku for helpful discussions. This work was supported by the National Science Foundation under the Grant No. ECCS-1408558.

\end{document}